\documentstyle[12pt]{article}
\begin{document}
\begin{flushright}
{\large PRL-TH/93-17 }
\vskip .2cm
{\large Rev. March 1994}
\end{flushright}

\begin{center}
{\Large \bf Weak dipole moment
of $\tau$ in $e^+e^-$ collisions\\[6pt] with longitudinally
polarized electrons}\\[50pt]
{\bf B. Ananthanarayan$^*$ and Saurabh D. Rindani}\\ \vskip .2in
Theory Group, Physical Research Laboratory\\Navrangpura,
Ahmedabad 380009, India\\[40pt]
{\bf \underline{Abstract}}\\[20pt]
\end{center}

It is pointed out that certain CP-odd momentum correlations in
the production and subsequent decay of tau pairs in $e^+e^-$
collisions get enhanced when the $e^-$ is longitudinally
polarized.  Analytic expressions for these correlations are
obtained for the single-pion decay mode of $\tau$ when
$\tau^+\tau^-$ have a  ``weak" dipole form factor (WDFF)
coupling to $Z$ .  For $e^+e^-$ collisions at the $Z$
peak, a sensitivity of about 1-5$\times 10^{-17}$\mbox{$e$ cm}
for the $\tau$ WDFF can be reached using a {\em single}
$\tau^+\tau^-$ decay channel, with $10^6\, Z$'s likely to be
available at the SLC at Stanford with $e^-$ polarization of
62\%-75\%.
\newpage

The standard model (SM), which has been well tested in recent
experiments, can adequately accommodate CP violation in the
K-meson system, the only place where it has been experimentally
observed. Nevertheless, there remains the intriguing possibility

that the underlying theory is an extension of SM which would
give observable CP violation in other hadronic, or even leptonic
systems.  For example, the observation of electric dipole
moments would signal physics beyond SM, since these are
predicted to be unobservably small in SM.  It is therefore
necessary to keep an eye open for such signals in the planning
and analysis of experiments.

The study of CP-odd correlations has been proposed in the past as
a test CP [1]. The observation of such correlations, for example
in $e^+e^-$ or $p\bar{p}$ experiments, would signal violation of
CP.  In particular, CP-odd observables arising due to a possible
$\tau$-lepton electric or ``weak" dipole moment have been analyzed
in great detail [2,3].  Such dipole moments can arise in
extensions of SM with CP violation coming from sectors other
than the standard $3\times 3$ Cabibbo-Kobayashi-Maskawa matrix.

Recent experiments at the LEP collider in CERN  have also put an
experimental upper limit [4] on the weak dipole form factor
(WDFF) of $\tau$ at the $Z$ resonance by looking for the tensor
correlation $T_{ij} = ({\bf q}_+ -
{\bf q}_-)_i ( {\bf q}_+\times
{\bf q}_- )_j$,
where ${\bf q}_+({\bf q}_-)$
represents the momentum of a charged particle arising from the
decay of $\tau^+(\tau^-)$ produced in $Z$ decay.

Recently, the Stanford Linear Collider (SLC) has achieved a
longitudinal polarization $(P_e)$ of 62\% for the electron beam,
which is likely to be increased to about 75\% [5].  Longitudinal
polarization of $e^-$ of 22\% has already been used for the
determination of $\sin ^2\theta_W$  using left-right asymmetry
[6]. Experiments with the present polarization of 62\% can lead
to better accuracy than obtained so far at LEP [7].

We study here the effect of longitudinal $e^-$ polarization on
CP-odd momentum correlations in $e^+e^- \rightarrow
\tau^+\tau^-$ with the subsequent decay of $\tau^+$ and
$\tau^-$. Since we have mainly the ongoing experiments at SLC in
mind, we concentrate on the $e^+e^-$ centre-of-mass (c.m.)
energy tuned to the $Z$ resonance. The CP-odd momentum
correlations [8] are associated with the c.m. momenta
${\bf p}$ of $e^-$,
${\bf q_+}$ of $\pi^+$ and
${\bf q_-}$ of $\pi^-$, where the $\pi^+$ and
$\pi^-$ arise in the decays $\tau^+ \rightarrow \pi^+
\bar{\nu}_{\tau}$
and $\tau^-\rightarrow \pi^-\nu_{\tau}$.  We restrict ourselves
to the single-pion decay mode of the tau, for which case we can
obtain analytic expressions for the correlations.

It must be noted that correlations which are CP violating in the
absence of initial beam polarization are not strictly CP odd
for arbitrary $e^+$ and $e^-$ polarizations, since the initial
state is then not necessarily  CP even. We argue, however,
that this is true to a high degree of accuracy in the case at
hand.

Our main result is that certain CP-odd correlations, which are
relatively small in the absence of $e^-$ polarization, since
they come with a factor $r=2g_{Ve}g_{Ae}/(g^2_{Ve} +
g^{2}_{Ae})$ ($\approx$ 0.16), get enhanced in the presence of
polarization, now being proportional to
$(r - P_e)/(1-rP_e)$ ($\approx
0.71$ for $P_e = -.62$) [9].  Here $g_{Ve}, g_{Ae}$ are the
vector and axial vector couplings of $e^-$ to $Z$.  The
correlations which have this property are those which have an
odd number of factors of the $e^+$ c.m. momentum
${\bf p}$, since this would need P and C
violation at the electron vertex.  Furthermore, we suggest a
procedure for  obtaining these correlations from the difference
in the event distributions for a certain polarization $P_e$ and
the sign-flipped polarization $-P_e$. With this procedure, the
correlations are further enhanced, leading to increased
sensitivity.

More specifically, we have considered the observables $O_1
\equiv \hat{\bf p}\cdot
\left({\bf q}_+\times {\bf q}_-\right)$
and $O_2\equiv\hat{\bf p}\cdot\left({\bf q}_++
{\bf q}_-\right)$ (the caret denoting a unit vector) and
obtained analytic expressions for their mean values and standard
deviations in the presence of longitudinal $e^-$ polarization
$P_e$.  By the procedure outlined above, the
magnitude of the WDFF at the $Z$, $\tilde{d}_{\tau}(m_Z)$
which can be measured at $1\sigma$ level is about 1-2$\times
10^{-16} e$ cm (3-5$\times 10^{-17} e$ cm) for a sample of 50,000
$(10^6)$ Z's, using a single $\tau$ decay channel.  Moreover, $O_2$,
being CPT-odd, measures ${\rm Im}\,\tilde{d}_{\tau}$, whereas $O_1$
measures ${\rm Re}\,\tilde{d}_{\tau}$.
Inclusion of other exclusive $\tau$
decay modes (not studied here) would improve the sensitivity
further.

The process we consider is
\begin{equation}
e^-(p_-) + e^+(p_+)\rightarrow \tau^-(k_-) + \tau^+(k_+),
\end{equation}
with the subsequent decays
\begin{equation}
\tau^-(k_-)\rightarrow \pi^-(q_-) + \nu_{\tau} ,\;
\tau^+(k_+)\rightarrow \pi^+(q_+) +\overline{\nu}_{\tau} .
\end{equation}

Under CP, the various three-momenta transform as
\begin{equation}
{\bf p}_-\leftrightarrow -
{\bf p}_+ ,\;
{\bf k}_-\leftrightarrow -
{\bf k}_+ ,\;  {\bf q}_-
\leftrightarrow - {\bf q}_+ .
\end{equation}
In the $e^+e^-$ c.m. frame, the only CP-odd vectors which can be
constructed out of the directly observable three-momenta ${\bf
p}\equiv -{\bf p}_-={\bf p}_+$, ${\bf q}_+$ and ${\bf q}_-$ , are
${\bf q}_+ + {\bf q}_-$ and ${\bf q}_+\times {\bf q}_-$ ,
while ${\bf p}$ and ${\bf q}_+ - {\bf q}_-$ are
the CP-even ones.  It is therefore simple to list possible
scalars which are odd under CP : ${\bf p}\cdot
\left({\bf q}_+ +{\bf q}_-\right) ,{\bf p}\cdot \left({\bf q}_+
\times {\bf q}_-\right)$ and $\left({\bf q}_+ +{\bf q}_-\right)\cdot
\left({\bf q}_+ -{\bf q}_-\right)$.  The CP-even scalars are
$\left({\bf q}_+ + {\bf q}_-\right)^2 , \left({\bf q_+} -
{\bf q_-} \right)^2$ and ${\bf p}\cdot \left({\bf q}_+ -
{\bf q}_- \right)$.  One can, of course, take
products among these six to  construct more CP-odd quantities.

We choose for our analysis the two CP-odd observables $O_1
\equiv \hat{\bf p}\cdot \left({\bf q}_+\times
{\bf q}_- \right)$ and $O_2 \equiv \hat{\bf p}\cdot
\left({\bf q}_+ + {\bf q}_-
\right)$, which have an odd number of factors of $\hat{\bf p}$.  As
mentioned before, they are expected to get enhanced in the
presence of $e^-$ polarization.
Though these observables are CP
odd, their observation with polarized $e^+$ and $e^-$ beams is
not necessarily an indication of CP violation, unless the $e^+$
and $e^-$ longitudinal polarizations are equal and opposite, so
that the initial state is described by a CP-even density matrix.
However, in the limit of $m_e=0$, the couplings of like-helicity
$e^+e^-$ pairs to spin-1 states like $\gamma$ and $Z$ drop out,
effectively giving rise to a CP-even initial state to a very good
accuracy for arbitrary $e^+$ and $e^-$ polarizations [10].

Of $O_1$ and  $O_2$, $O_1$ is even under the combined CPT
transformation, and $O_2$ is CPT-odd. A CPT-odd
observable can only have a non-zero value in the presence of an
absorptive part of the amplitude.  It is therefore expected that
$\langle O_2\rangle $ will be proportional to the imaginary part
of the weak
dipole form factor ${\rm Im}\,\tilde{d}_{\tau}$ , since final-state
interaction, which could give rise to an absorptive part,
is negligible in the weak $\tau$ decays. Since $\langle
O_1\rangle $ and mean
values of other CPT-even quantities will be proportional to
${\rm Re}\,\tilde{d}_{\tau}$, phase information on
$\tilde{d}_{\tau}$ can only be obtained if $\langle O_2\rangle $ (or
some other CPT-odd quantity) is also measured.

We assume SM couplings for all particles except $\tau$, for which
an additional WDFF interaction is assumed, viz.,
\begin{equation}
{\cal L}_{WDFF} = - \frac{i}{2} \tilde{d}_{\tau}
\overline{\tau}\sigma^{\mu\nu}\gamma_5\tau
\left(\partial_{\mu}Z_{\nu} - \partial_{\nu}Z_\mu\right) ,
\end{equation}
where $\tilde{d}_{\tau} \equiv \tilde{d}_{\tau}
(s\!=\!m^2_Z)$.  Using (4), we now proceed to calculate
$\langle O_1\rangle $ and $\langle O_2\rangle $ in the presence
of longitudial polarization $P_e$ for $e^-$.

We can anticipate the effect of $P_e$ in general for the process
(1).  We can write the matrix element squared for the process in
the leading order in perturbation theory, neglecting the
electron mass, as
\begin{equation}
\vert M\vert^2 = \sum_{i,j} L^{ij}_{\mu\nu}(e) L^{ij
\mu\nu*}(\tau) \frac{1}{s-M^2_i}\, \frac{1}{s-M^2_j} ,
\end{equation}
where the summation is over the gauge bosons $(\gamma ,Z,\ldots)$
exchanged in the $s$ channel, and $L^{ij}_{\mu\nu}(e,\tau)$
represent the tensors arising at the $e$ and $\tau$ vertices:
\begin{equation}
L^{ij}_{\mu\nu} = V^i_\mu V^{j*}_{\nu} .
\end{equation}
For the electron vertex, with only the SM  vector and axial vector
couplings,
\begin{equation}
V^i_{\mu}(e) = g_i\bar{v}(p_+ ,
s_+)\gamma_{\mu}\left(g^i_{Ve}-\gamma_5g^i_{Ae}\right)u(p_- ,
s_-),
\end{equation}
$g_i$ being the appropriate coupling constant $\left( g_{\gamma}
= e, g_Z = g/(2\cos\theta_W)\right)$, and $g^i_{Ve}$ and
$g^i_{Ae}$ are given by
\begin{equation}
g^{\gamma}_{Ve} = - 1,\; g^{\gamma}_{Ae} = 0 ;
\end{equation}
\begin{equation}
g^Z_{Ve} = -\frac{1}{2} + 2\sin ^2\theta_W ,\; g^Z_{Ae} =
-\frac{1}{2} .
\end{equation}
It is easy to check, by putting in helicity projection operators,
that
\begin{eqnarray}
\lefteqn {L^{ij}_{\mu\nu}(e)=g_ig_j} &&\nonumber \\
 &\! \times&\!\!\!\! \left\{\left[\left( 1-P_eP_{\bar
e}\right) \left(g^i_{Ve}g^j_{Ve}
+ g^i_{Ae}g^j_{Ae}\right) - \left( P_e-P_{\bar
e}\right)\left(g^i_{Ve}g^j_{Ae} +
g^i_{Ae}g^j_{Ve}\right) \right]
Tr(\rlap{$p$}/_-\gamma_{\mu}\rlap{$p$}/_+\gamma_\nu) \right.
\nonumber \\
 &\! +&\!\!\!\! \left.  \left[ \left( P_e-P_{\bar e}\right)
\left(g^i_{Ve}g^j_{Ve} + g^i_{Ae}g^j_{Ae}\right) -
\left( 1-P_eP_{\bar e}\right)\left(g^i_{Ve}g^j_{Ae} +
g^i_{Ae}g^j_{Ve}\right)\right]
Tr \left(\gamma_5\rlap{$p$}/_-
\gamma_{\mu}\rlap{$p$}/_+\gamma_\nu \right) \right\} \nonumber\\
&&
\end{eqnarray}
in the limit of vanishing electron mass, where $P_e$ ($P_{\bar
e}$) is the
degree of the $e^-$ ($e^+$) longitudinal polarization. Note that
the combinations $P_e-P_{\bar e}$ and $1-P_eP_{\bar e}$ occurring
in (10) are indeed CP-even, showing that the initial state is
effectively CP even for arbitrary $P_e$, $P_{\bar e}$.

Eq.(10) gives a simple way of incorporating the effect of the
longitudinal polarization.  In particular, at the $Z$ peak, where
photon effects can be neglected, to go from the unpolarized to
the polarized one, one has to make the replacement (henceforth,
we drop the superscript $Z$, as we shall only deal with $Z$
couplings):
\begin{equation}
\begin{array}{clclc}
g^2_{Ve} + g^2_{Ae} &\rightarrow &g^2_{Ve} + g^2_{Ae} &-& P_e \,2
g_{Ve} g_{Ae},  \\
2g_{Ve}g_{Ae} &\rightarrow &2g_{Ve}g_{Ae} &-
&P_e \left(g^2_{Ve} + g^2_{Ae}\right).
\end{array}
\end{equation}
We have set $P_{\bar e}=0$, as is the case at SLC, for example.
It is then clear from (9), using $\sin^2\theta_W \approx 0.23$,
that quantities which are suppressed in the absence of
polarization because of the small numerical value of
$2g_{Ve}g_{Ae}$ will get considerably enhanced in the presence
of polarization.

To calculate correlations of $O_1$ and $O_2$, we need the
differential cross section for (1) followed by (2) at the $Z$
peak, arising from SM $Z$ couplings of $e$ and $\tau$, together
with a weak-dipole coupling of $\tau$ arising from eq.(4).  (We
neglect electromagnetic effects completely).  The calculation may be
conveniently done, following ref.[3] (see also ref.[11]), in
steps, by first determining the production matrix $\chi$ for
$\tau^+\tau^-$ in spin space, and then taking
its trace with the decay matrices $\cal{D}^{\pm}$ for $\tau^{\pm}$
decays into single pions:

\begin{equation}
\frac{1}{\sigma}\,
\frac{d\sigma}{d\Omega_kd\Omega^*_-d\Omega^*_+dE^*_-dE^*_+}
= \frac{k}{8\pi m^2_Z\Gamma(Z\rightarrow\tau^+\tau^-)}
\frac{1}{(4\pi)^3}
\chi^{\beta\beta^{\prime}, \alpha\alpha^{\prime}}
{\cal D}^-_{\alpha^\prime\alpha} {\cal D}^+_{\beta^\prime\beta},
\end{equation}
where $d\Omega_k$ is the solid angle element for
${\bf k}_+$ in the overall c.m. frame, $k =
\vert{\bf k}_+\vert$, and $d\Omega^*_{\pm}$ are
the solid angle elements for ${\bf q}^*_{\pm}$,
the $\pi^{\pm}$ momenta in the $\tau^{\pm}$ rest frame.  The
${\cal D}$ matrices are given by
\begin{equation}
{\cal D}^{\pm} = \delta\left(E^*_{\pm} - E_0\right)\left[ 1 \mp
{\bf \sigma}_{\pm} \cdot\hat{\bf q}^*_{\pm} \right],
\end{equation}
where ${\bf \sigma}_{\pm}$ are the Pauli
matrices corresponding to the $\tau^{\pm}$ spin, $E^*_{\pm}$ are
the $\pi^{\pm}$ energies in the $\tau^{\pm}$ rest frame, and
\begin{equation}
E_0 = \frac{1}{2} m_{\tau} (1 + p) ;\; p = m^2_{\pi}/m^2_{\tau}.
\end{equation}

The expressions for $\chi$ arising from SM as well as WDFF
coupling of $\tau$ are rather long, and we refer the reader to
ref.[3] for these expressions in the absence of polarization.
It is straightforward to incorporate polarization using (11).

The expressions for the correlations $\langle O_1\rangle $ and
$\langle O_2\rangle$ obtained
are, neglecting $\tilde{d}^2_{\tau}$,
\begin{equation}
\langle \hat{\bf p}\cdot\left({\bf q}_+\times
{\bf q}_-\right)\rangle = -
\frac{m^2_Z}{18e}m_{\tau}{\rm Re}\,\tilde{d}_{\tau}
c_Ws_W(1-x^2)\left(\frac{r - P_e}{1-rP_e}\right)
\frac{g_{A\tau}(1-p)^2 -
3g_{V\tau}(1-p^2)}{g^2_{V\tau} (1+\frac{1}{2}x^2) +
g^2_{A \tau}
\left( 1-x^2\right)} ,
\end{equation}
and
\begin{equation}
\langle \hat{\bf p}\cdot({\bf q}_+ +
{\bf q}_-)\rangle = \frac{2m_Z}{3e}
m_{\tau}{\rm Im}\,\tilde{d}_{\tau}c_Ws_W
\left(\frac{r - P_e}{1-rP_e}\right)
\frac{g_{A\tau}(1-x^2)^2
(1-p)}{g^2_{V\tau}(1+\frac{1}{2}x^2) + g^2_{A\tau} (1-x^2)},
\end{equation}
where $c_W=\cos\theta_W$, $s_W=\sin\theta_W$, $x =
2m_{\tau}/m_Z$, and $r=2g_{Ve}g_{Ae}/(g^2_{Ve}+g^2_{Ae})$.

We have also obtained analytic expressions for the variance
$\langle  O^2\rangle  - \langle  O\rangle ^2 \approx \langle
O^2\rangle $  in each case, arising from the CP-invariant SM
part of the interaction, but since they are lengthy, they will
be presented elsewhere [12].

Our numerical results are presented in the tables.  In Tables 1
and 2 we have presented, as in  ref.[3], the values of
$c_{\pi\pi}$ for $O_1$ and $O_2$ respectively, defined as
 the correlation for a value of ${\rm Re}\,\tilde{d}_{\tau}$ or
${\rm Im}\,\tilde{d}_{\tau}$ (as the case may be) equal to
$e/m_Z$, for various values of $P_e$. (By our convention, a
positive value of $P_e$ means right-handed polarization, and
vice-versa).
We have also presented the value of $\sqrt{\langle O^2\rangle }$ and
$\delta\tilde{d}_{\tau}$, which represents the 1 s.d.
upper limit unit on $\tilde{d}_{\tau}$ which can be
placed with a certain sample of events, for 50,000 $Z$'s currently
seen at SLC with 62\% polarization, and for $10^6\, Z$'s,
eventually hoped to be achieved. This 1 s.d. limit is the value
of $\tilde{d}_{\tau}$ which gives a mean value of $O_i$
equal to the s.d. $\sqrt{\langle O^2_i\rangle /N_{\pi\pi}}$ in
each case.
\begin{center}
\begin{tabular}{||c|c|c|c|c||}
\hline
$P_e$ &$c_{\pi\pi}$ &
$\sqrt{\langle  O^2_1\rangle }$ &\multicolumn{2}{|c||}{$\delta
{\rm Re}\,\tilde{d}_{\tau}(e$ cm) for}\\
\cline{4-5}
 &(GeV$^2$) &(GeV$^2$)&
{$5\times 10^4\, Z$'s} &{$10^6
\, Z$'s}\\
\hline
0 &0.898 &$12.861$ &$6.6\times 10^{-16}$ &$1.5\times 10^{-16}$\\
$+0.62$ &$-2.890$ &12.861 &$2.0\times 10^{-16}$ &$4.6\times 10^{-17}$\\
$-0.62$ &$4.007$ &12.861 &$1.5\times 10^{-16}$ &$3.3\times 10^{-17}$\\
$+0.75$ &$-3.792$ &12.861 &$1.6\times 10^{-16}$ &$3.5\times 10^{-17}$\\
$-0.75$ &$4.589$ &12.861 &$1.3\times 10^{-16}$ &$2.9\times 10^{-17}$\\
\hline
\end{tabular}
\\ \vskip .2cm
{Table 1: Results for the observable $O_1 \equiv \hat{\bf p}\cdot
({\bf q}_+\times {\bf q}_-)$}
\end{center}

\begin{center}
\begin{tabular}{||c|c|c|c|c||}
\hline
$P_e$ &$c_{\pi\pi}$
&$\sqrt{\langle  O^2_1\rangle }$&\multicolumn{2}{c||}{$\delta
{\rm Im}\,\tilde{d}_{\tau} (e$ cm) for} \\
\cline{4-5}
&(GeV) &(GeV)
&$5\times 10^4\, Z$'s
&$10^6\,  Z$'s\\
\hline
0 &$-0.157$ &9.572 &$2.8\times 10^{-15}$ &$6.2\times 10^{-16}$\\
$+0.62$ &$0.505$ &9.572 &$8.7\times 10^{-16}$ &$1.9\times 10^{-16}$\\
$-0.62$ &$-0.700$ &9.572 &$6.3\times 10^{-16}$ &$1.4\times 10^{-16}$\\
$+0.75$ &$0.662$ &9.572 &$6.6\times 10^{-16}$ &$1.5\times 10^{-16}$\\
$-0.75$ &$-0.802$ &9.572 &$5.5\times 10^{-16}$ &$1.2\times 10^{-16}$\\
\hline
\end{tabular}
\\ \vskip .2cm
{Table 2: Results for the observable $O_2 \equiv \hat{\bf p}\cdot
({\bf q}_++ {\bf q}_-)$}
\end{center}

As can be seen from Tables 1 and 2, $\langle O_1\rangle $ and
$\langle O_2\rangle $ can probe
respectively ${\rm Re}\,\tilde{d}_{\tau}$ and
${\rm Im}\,\tilde{d}_{\tau}$ down to about $2\times 10^{-16} e$ cm
and $6\times 10^{-16} e$ cm with the current data from SLC.  These
limits can be improved by a factor of 5 in future runs of SLC.

These limits can be considerably improved by looking at
correlations of the same observables, but in a sample obtained
by counting the difference between the number of events for a certain
polarization, and for the corresponding sign-flipped polarization.
If the differential cross section for the process for an
$e^-$ polarization $P_e$ is given by
\begin{equation}
\frac{d\sigma}{d^3q_-d^3q_+}(P_e)=A({\bf p},{\bf q}_-,{\bf q}_+) +
P_e\cdot B({\bf p},{\bf q}_-,{\bf q}_+),
\end{equation}
then one calculates average values over the distribution given
by the difference
\begin{equation}
\frac{d\sigma (P_e)}{d^3q_-d^3q_+} -
\frac{d\sigma (-P_e)}{d^3q_-d^3q_+}
 =2 P_e B({\bf p},{\bf q}_-,{\bf q}_+).
\end{equation}
The correlations are considerably larger, whereas the variances
are unchanged. The event sample is however smaller (by a factor
of about .16$P_e$, which is the left-right asymmetry times degree
of polarization), leading to a larger statistical error. The
sensitivity is nevertheless improved, as can be seen from Table
3. There we present, for the two variables $O_1$ and $O_2$,
$c_{\pi\pi}$ (defined as before) for the polarization asymmetrized
sample described above,
and the corresponding  quantities $\sqrt{\langle  O_i^2 \rangle }$
and the 1 s.d. limits on ${\rm Re}\,\tilde{d}_\tau$ and
${\rm Im}\,\tilde{d}_\tau$, respectively, both called
$\delta\tilde{d}_\tau$ in this table for convenience. We give
the limits for two
luminosities, but only for one polarization, $P_e=.62$. Only the
last column changes with $P_e$, and the
improvement in going to $P_e=.75$ is marginal.
\begin{center}
\begin{tabular}{||c|c|c|c|c||}
\hline
 &$c_{\pi\pi}$ &$\sqrt{\langle  O^2_i\rangle }$
&\multicolumn{2}{c||}{$\delta
\tilde{d}_\tau (e\;{\rm cm})$ for} \\
\cline{4-5}
& & &$5\times 10^4\, Z$'s
&$10^6\,  Z$'s\\
\hline
$O_1$ &35.545 GeV$^2$&12.861 GeV$^2$ &$5.3\times 10^{-17}$ &
$1.2\times 10^{-17}$\\
$O_2$ &$-6.208$ GeV&$9.572$ GeV &$2.2\times 10^{-16}$
&$5.0\times 10^{-17}$\\
\hline
\end{tabular}
\\ \vskip .2cm
Table 3: Results for the
polarization asymmetrized sample of events, for $P_e=0.62$.
\end{center}

 As seen from Table 3, the current
data can yield upper limits of about 5-$22\times 10^{-17} e$ cm for
${\rm Re}\,\tilde{d}_{\tau}$ or ${\rm Im}\,\tilde{d}_{\tau}$. An
integrated luminosity corresponding to $10^6$ Z's
at SLC can improve the limits to about 1-$5\times 10^{-17} e$ cm.
These should be compared with the 95\% C.L.
limit of ${\rm Re}\,\tilde{d}_{\tau}$ $< 7.0\times 10^{-17} e$
cm from OPAL and
${\rm Re}\,\tilde{d}_{\tau}$ $< 3.7\times 10^{-17} e$ cm from
ALEPH obtained by looking for tensor correlations from a sample
of about 650,000 Z's at LEP, and including several decay
channels of $\tau$ [4].
The real advantage of polarization can be seen in the
sensitivity for the measurement of ${\rm Im}\tilde{d}_{\tau}$.
The limit obtainable at LEP under ideal
experimental conditions from the 2$\pi$ channel using the tensor
correlation $\langle
\hat{\bf p}\cdot (\hat{\bf q}_++\hat{\bf q}_-)\hat{\bf p}\cdot
(\hat{\bf q}_+ -\hat{\bf q}_- )\rangle $ is $10^{-16}$ $e$ cm
with $10^7$ $Z$'s [3], whereas the limit we obtain here in the
presence of polarization is as low as
$5\times 10^{-17}$ $e$ cm with only $10^6$ $Z$'s.

To conclude, our results show that the CP-odd correlations
$\langle O_1\rangle $
and $\langle O_2\rangle $ are considerably enhanced due to
longitudinal $e^-$
polarization.  By considering  values of these
correlations on changing the sign of the polarization, a much
greater sensitivity is obtained. Considering both these
correlations together can give phase information on the WDFF,
which cannot be obtained using any single correlation.

We must emphasize that we have considered here only one $\tau$
decay channel.  Since the
number of events in any single channel is small, inclusion of
additional decay modes is necessary and can improve the
sensitivity.
 The sensitivity of other channels will be
discussed in a future work [12]. The limits obtainable may be
comparable  to those from tensor correlations [3,4] with
unpolarized beams, as used at LEP. Moreover, studying $O_2$ can
constrain ${\rm
Im}\,\tilde{d}_\tau$, the tensor correlation could only
constrain ${\rm Re}\,\tilde{d}_\tau$. It should also be noted
that the tensor
correlations considered in refs. [3,4] are not changed in the
presence of polarization.

 We have neglected errors in measurement of $P_e$,
whose effect on the results would be small compared to that of
statistical errors.

It is possible that longitudinal polarization may be available
at tau-charm factories or other future linear colliders planned
to operate at higher
energies.  The advantages of polarization presented here can be
studied by extending the analysis to the relevant energies,
including the effect of photon exchange and the $\tau$ electric
dipole coupling.  The study could also be extended to dipole
moments of the top-quark
or $W^{\pm}$  which could be pair produced at the
high energy accelerators.

It would also be interesting to consider event asymmetries
corresponding to $O_1$ and $O_2$ (rather than their mean values), and
also other CP-odd
correlations with an odd number of $\hat{\bf p}$'s, like
$\langle  ({\bf q}^2_+ -
{\bf q}^2_- )
\hat{\bf p}\cdot({\bf q}_+ -
{\bf q}_-)\rangle $, which may have a different
sensitivity.
\vskip .2cm
\centerline{\bf Acknowledgements}

We thank Rohini Godbole, Anjan Joshipura and Marek Nowakowski
for useful
discussions and suggestions, and for a critical reading of the
manuscript. We also thank D. Indumathi,
B.R. Sitaram and Sridhar K. for helpful discussions. S.D.R.
thanks Prof. L.M. Sehgal for drawing his attention to the
preprint quoted in [10].
\vskip 1cm
\newpage
\begin{center}
{\large \bf References and Footnotes}
\end{center}

\noindent$^*$ Present address: Institut de Physique Th\'eorique,
B\^atiment des Sciences Physiques, Universit\'e de Lausanne, CH
1015, Lausanne, Switzerland.
\vskip .2cm

\noindent [1]  J.F. Donoghue and G. Valencia, Phys. Rev. Lett. {\bf 58}, 451
(1987); M. Nowakowski and A. Pilaftsis, Mod. Phys. Lett. A {\bf
4}, 829 (1989), Z. Phys. C {\bf 42}, 449
(1989); W. Bernreuther, U. L\" ow, J.P. Ma and O. Nachtmann,
Z. Phys. C {\bf 43}, 117 (1989); W.
Bernreuther and O. Nachtmann, Phys. Lett. B {\bf 268}, 424
(1991); M.B. Gavela {\it et al.}, Phys. Rev. D {\bf 39}, 1870 (1989).

\noindent [2] F. Hoogeveen and L. Stodolsky, Phys. Lett. B {\bf 212},
505 (1988); S. Goozovat and C.A. Nelson, Phys. Lett. B {\bf
267}, 128 (1991); G. Couture, Phys. Lett. B {\bf 272}, 404
(1991); W. Bernreuther and O. Nachtmann, Phys. Rev. Lett. {\bf
63}, 2787 (1989).

\noindent [3] W. Bernreuther, G.W. Botz, O. Nachtmann and P. Overmann,
Z. Phys. C {\bf 52}, 567 (1991); W.
Bernreuther, O. Nachtmann and P. Overmann, Phys. Rev. D {\bf
48}, 78 (1993).

\noindent [4] OPAL Collaboration, P.D. Acton {\it et al.}, Phys.
Lett. B {\bf
281}, 405 (1992); ALEPH Collaboration, D. Buskulic {\it et al.}, Phys.
Lett. B {\bf 297}, 459 (1992).

\noindent [5] M. Swartz, private communication.

\noindent [6] SLD Collaboration, K. Abe {\it et al.}, Phys. Rev. Lett. {\bf
70}, 2515 (1993).

\noindent [7] M.J. Fero, SLAC report SLAC-PUB-6027 (1992), to be published.

\noindent [8] The momentum correlations arise due to $\tau$ polarization,
and may be thought of as correlations among $\tau$ momenta and
spins, spin-analyzed by $\tau$ decay.  In practice, we need
consider only the directly observable momentum correlations.

\noindent [9] A similar enhancement for the observation of CP violation
in B decays was pointed out by W.B. Atwood, I. Dunietz and P.
Grosse-Wiesmann, Phys. Lett. B {\bf 216}, 227 (1989).

%
%

\noindent[10] The exact extent to which CP-invariant
interactions would contribute to the correlations would be model
dependent. However, it is clear that at the $Z$ resonance, the
corrections to zeroth order in $\alpha$ would be suppressed at
least by a factor $m_e/m_Z \approx 6\times 10^{-6}$. There are
however corrections at order $\alpha$ from tree-level helicity-flip
collinear photon emission which are not suppressed by electron
mass. The corresponding cross section being non-resonant is
suppressed by a factor of about $10^{-4}$, (see B. Falk and L.M.
Sehgal, Aachen report no. 93/29 (1993), to appear in Phys. Lett. B)
and can be neglected. There can be non-resonant CP-odd  helicity
combinations which contribute at loop level. Since these do not
interfere with the leading $Z$ contributions coming from CP-even
helicity combinations, they will be suppressed by a factor
$\alpha ^2\Gamma^2_Z/m^2_Z  \approx 5\times 10^{-8}$.

\noindent [11] Y.S. Tsai, Phys. Rev. {\bf D4}, 2821 (1971); {\bf 13},
771 (1976) (E); S. Kawasaki, T. Shirafuji and Y.S. Tsai, Prog.
Theo. Phys. {\bf 49}, 1656 (1973).

\noindent [12] B. Ananthanarayan and S.D. Rindani, PRL Ahmedabad preprint
PRL-TH-94/7.

\end{document}